\documentclass[%
 reprint,
superscriptaddress,
 amsmath,amssymb,
aps,
prb,
]{revtex4-2}

\newcommand*{\affmanpa}{Department of Physics and Astronomy, University of Manchester, Manchester M13 9PL, United Kingdom}
\newcommand*{\affmancs}{Department of Computer Science, University of Manchester, Manchester M13 9PL, United Kingdom}
\newcommand*{\affconacyt}{Consejo Nacional de Ciencia y Tecnolog\'{i}a (CONACyT), Av. Insurgentes Sur 1582, Col. Cr\'{e}dito Constructor, Alcald\'{i}a Benito Juarez, C.P. 03940, Ciudad de M\'{e}xico, M\'{e}xico}
\newcommand*{\affec}{Facultad de Ciencias Naturales y Matem\'{a}ticas, Centro de Investigaci\'{o}n y Desarrollo en Nanotecnolog\'{i}a, Escuela Superior Polit\'{e}cnica del Litoral, ESPOL, Campus Gustavo Galindo, Km.~30.5 V\'{i}a Perimetral, Guayaquil, 090902, Ecuador}

\usepackage{graphicx}
\usepackage{dcolumn}
\usepackage{bm}
\usepackage{gensymb}
\usepackage{hyperref}

\usepackage[dvipsnames]{xcolor} 

\usepackage{blindtext}
\usepackage{outlines}
\usepackage{comment} 
\usepackage{siunitx} 

\begin{document}

\preprint{APS/123-QED}

\title{Oblique spin injection to graphene via geometry controlled magnetic nanowires}

\author{Jesus C.\ Toscano-Figueroa}
\altaffiliation{These authors contributed equally to the work.}
\affiliation{\affmanpa}
\affiliation{\affconacyt}

\author{Daniel Burrow}
\altaffiliation{These authors contributed equally to the work.}
\affiliation{\affmanpa}

\author{Victor H.\ Guarochico-Moreira}
\affiliation{\affmanpa}
\affiliation{\affec}

\author{Chengkun Xie}
\affiliation{\affmanpa}

\author{Thomas Thomson}
\affiliation{\affmancs}

\author{Irina V.\ Grigorieva}
\affiliation{\affmanpa}

\author{Ivan J.\ Vera-Marun}
\thanks{Correspondence to ivan.veramarun@manchester.ac.uk and daniel.burrow@manchester.ac.uk}
\affiliation{\affmanpa}

\date{\today}

\begin{abstract}
We exploit the geometry of magnetic nanowires, which define 1D contacts to an encapsulated graphene channel, to introduce an out-of-plane component in the polarisation of spin carriers. By design, the magnetic nanowires traverse the angled sides of the 2D material heterostructure. Consequently, the easy axis of the nanowires is inclined, and so the local magnetisation is oblique at the injection point. As a result, when performing non-local spin valve measurements we simultaneously observe both switching and spin precession phenomena, implying the spin population possesses both in-plane and out-of-plane polarisation components. By comparing the relative magnitudes of these components, we quantify the angle of the total spin polarisation vector. The extracted angle is consistent with the angle of the nanowire at the graphene interface, evidencing that the effect is a consequence of the device geometry. This simple method of spin-based vector magnetometry provides an alternative technique to define the spin polarisation in 2D spintronic devices.
\end{abstract}

\maketitle


\section{\label{sec:level1}Introduction}

Efficient generation of spin accumulation in an appropriate transport medium, and control over the consequent diffusive spin signals, are two of the most crucial elements of spintronic devices \cite{Datta-1990, Wolf-2001}. Graphene is an ideal material for spin transport, owing to its high electron mobility, low spin orbit coupling strength, and gate tuneable carrier density \cite{Novoselov-2004, CastroNeto-2009, Tombros-2007, Han-2011, Vera-Marun-2011}. Furthermore, encapsulating graphene between two layers of hexagonal-boron nitride (hBN), an insulating isotrope of graphite, is known to greatly increase its quality as both a charge and spin transport medium, with significant improvements to electron mobility and spin diffusion lengths reported \cite{Dean-2010, Zomer-2011, Guimaraes-2014, InglaAynes-2015, InglaAynes-2016, Gurram-2018}. Making a Van der Waals heterostructure comprising insulating outer layers, inherently presents challenges in contacting graphene, with conventional top contacts no longer a viable option. A promising alternative is the use of one-dimensional (1D) edge contacts, where electrodes make contact with the graphene channel only along its atomic edge \cite{Wang-2013a}. Such contacts have been shown to be far less invasive than top contacts \cite{Giovannetti-2008, Karpiak-2017, Choi-2022}, and have recently been implemented, in the case of magnetic nanowire electrodes, for efficient spin injection and long range spin transport \cite{Guarochico-Moreira-2022}. 

\begin{figure*}
\includegraphics[width=\textwidth]{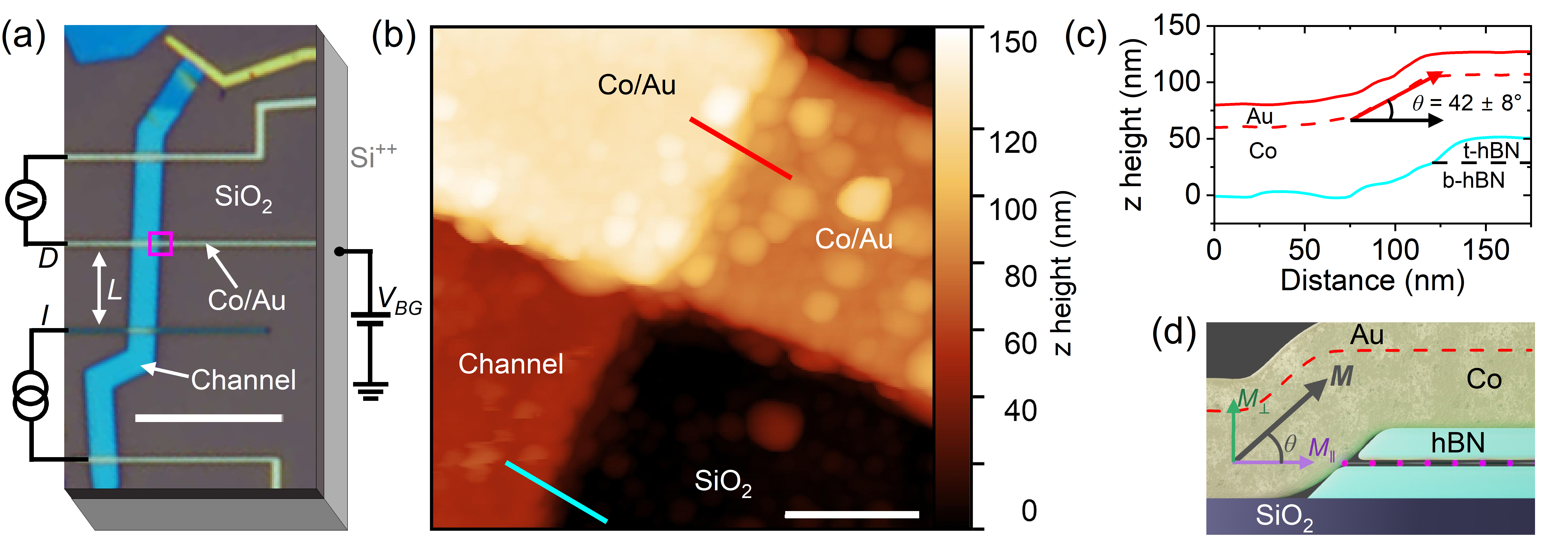}
\caption{Characterisation of device A. \textbf{(a)} Optical image of device. The blue strip is the hBN graphene encapsulated channel, while the yellow bars are Co/Au nanowires. Scale bar is 10 $\mu$m. The non-local measurement configuration is shown, with the injector ($I$) and detector ($D$) electrodes, as well as the back gate connection ($V_{BG}$), indicated. $L$ is the channel length in the non-local measurement. \textbf{(b)} AFM scan of a magnetic nanowire, taken from the region enclosed by the box in panel (a). Cyan and red bars indicate the location of profiles taken from the scan. Scale bar is 200 nm. \textbf{(c)} Height profiles of the stack and magnetic nanowire, taken from panel (b). The angle of inclination of the nanowire is found to be $42\pm8 \degree$. \textbf{(d)} Schematic illustration of the device profile, showing the hBN/graphene/hBN heterostructure with a Co/Au magnetic nanowire laid over the top. The angled magnetisation at the injection point is indicated.}
\label{fig:DevSchematic}
\end{figure*}

Here, we investigate how the geometry of these magnetic nanowires can be exploited to define the polarisation of a spin population injected into a high quality graphene device \cite{Gentile-2022}. The magnetic nanowires are engineered with considerable shape anisotropy (Figure \ref{fig:DevSchematic}a), such that the easy axis points along their length. Moreover, as the nanowires cross the 2D material heterostructure the easy axis tilts out-of-plane at the interface with the graphene (Figure \ref{fig:DevSchematic}d). As such, the local magnetisation vector is oblique at the point of spin injection; in other words, it is described by both in-plane and out-of-plane components (Figure \ref{fig:DevSchematic}d). To explore the effect of the magnetisation angle on diffusive spin transport through the graphene channel, we conduct standard spin valve (magnetic field in-plane) and Hanle spin precession (magnetic field out-of-plane) measurements \cite{Zutic-2004, Tombros-2007, Popinciuc-2009}. We find the spin valve response displays a Hanle-like line shape, implying that the oblique magnetisation transmits an out-of-plane component to the polarisation of the injected spin population, which precesses around the in-plane magnetic field. By analysing the precession curve observed in our spin valve measurements, and comparing the results to those extracted from standard Hanle spin precession data, we quantify the relative magnitudes of in-plane and out-of-plane spin polarisation and use this information to reconstruct the angle of the total polarisation vector. The extracted angle is consistent with the inclination of the magnetic nanowire at the graphene interface, evidencing that the effect is geometric in nature. Previous demonstrations of generation of out-of-plane injected spin polarisation in lateral devices, have utilised complex physics such as the anomalous spin Hall effect \cite{Das-2018}, magnetocrystalline anisotropy within Van der Waals magnets \cite{Zhao-2023}, or generation within graphene itself via proximity-induced spin-orbit coupling \cite{safeer_spin_2020}.
Hence, exploiting device geometry provides an alternative, and relatively simple, technique for defining spin polarisation in spintronic devices.

\section{\label{sec:frabication_and_methods}Fabrication and characterisation}

We fabricate a total of four devices (A-D) in line with the methods used in our previous work \cite{Guarochico-Moreira-2022}. Individual flakes of 2D material (both hBN and graphene) are isolated via mechanical exfoliation, then subsequently stacked into hBN/graphene/hBN heterostructures via the dry peel transfer technique \cite{Dean-2010, Mayorov-2011}. Reactive ion etching is used, along with a polymer hard mask, to engineer a long, narrow transport channel out of the heterostructure (Figure \ref{fig:DevSchematic}a). After etching, the sides of the stack are angled and a small ledge of graphene is left exposed at the edges (Figure \ref{fig:DevSchematic}b,c). This permits one-dimensional electrical contact to the heterostructure by deposition of ferromagnetic metal nanowires \cite{Wang2013, Karpiak-2017}. The nanowire pattern is defined in a polymer layer via electron beam lithography, before metal is deposited via electron beam evaporation; materials used for all devices are 60 nm Cobalt (Co) with a 20 nm Gold (Au) capping layer. 

Devices fabricated using this methodology have previously shown high electronic quality, with charge carrier mobilities $>10$ m$^2$ V$^{-1}$ s$^{-1}$, and demonstrated efficient spin transport with spin relaxation lengths $>10$ $\mu$m \cite{Guarochico-Moreira-2022}. One aspect of this architecture that remains unexplored, is the geometry of the magnetic nanowires that form the 1D contacts. The nanowire design results in an easy axis dictated by shape anisotropy, which points along the length of the wire. Therefore, switching of the in-plane magnetisation, $M_{||}$, can be achieved with relatively low magnetic fields, as expected for Co nanowires with dominant shape anisotropy \cite{Henry-2001}. However, owing to the geometry of the nanowires as they traverse the sides of heterostructure, the easy axis is inclined and the magnetisation gains a small out-of-plane component, $M_{\perp}$, at the injection point (Figure \ref{fig:DevSchematic}d). 

A further consideration in this geometric treatment, is the role of the heterostructure thickness, $t$, which is determined entirely by the encapsulating hBN layers. The thicknesses of our four heterostructures are varied by selection of hBN flakes (thickness decreases from A-D), while the thickness of the Co layer in the nanowires is kept constant across all devices (60 nm). Figures \ref{fig:DevSchematic}a and \ref{fig:DevSchematic}b show optical and AFM images of device A, respectively. Profiles taken from the AFM scan show the total height of this heterostructure to be $\sim 50$ nm (Figure \ref{fig:DevSchematic}c), which is comparable to the thickness of the Co within the nanowire $\sim 60$ nm. Aside from protecting the graphene from environmental contamination and giving rise to high charge mobilities, using thicker flakes of hBN ($>20$ nm) is expected to lead to a greater magnitude of $M_\perp$ in the nanowires, relative to a given thickness of Co. It should be noted that the etching process used has to be optimised for thicker flakes. Using AFM profile data for device A, the angle of inclination of the nanowires is found to be $\theta = 42\pm 8 \degree$. 

\section{\label{sec:Oblique_spin_injection}Oblique spin injection}

\begin{figure*}
\includegraphics[width=\textwidth]{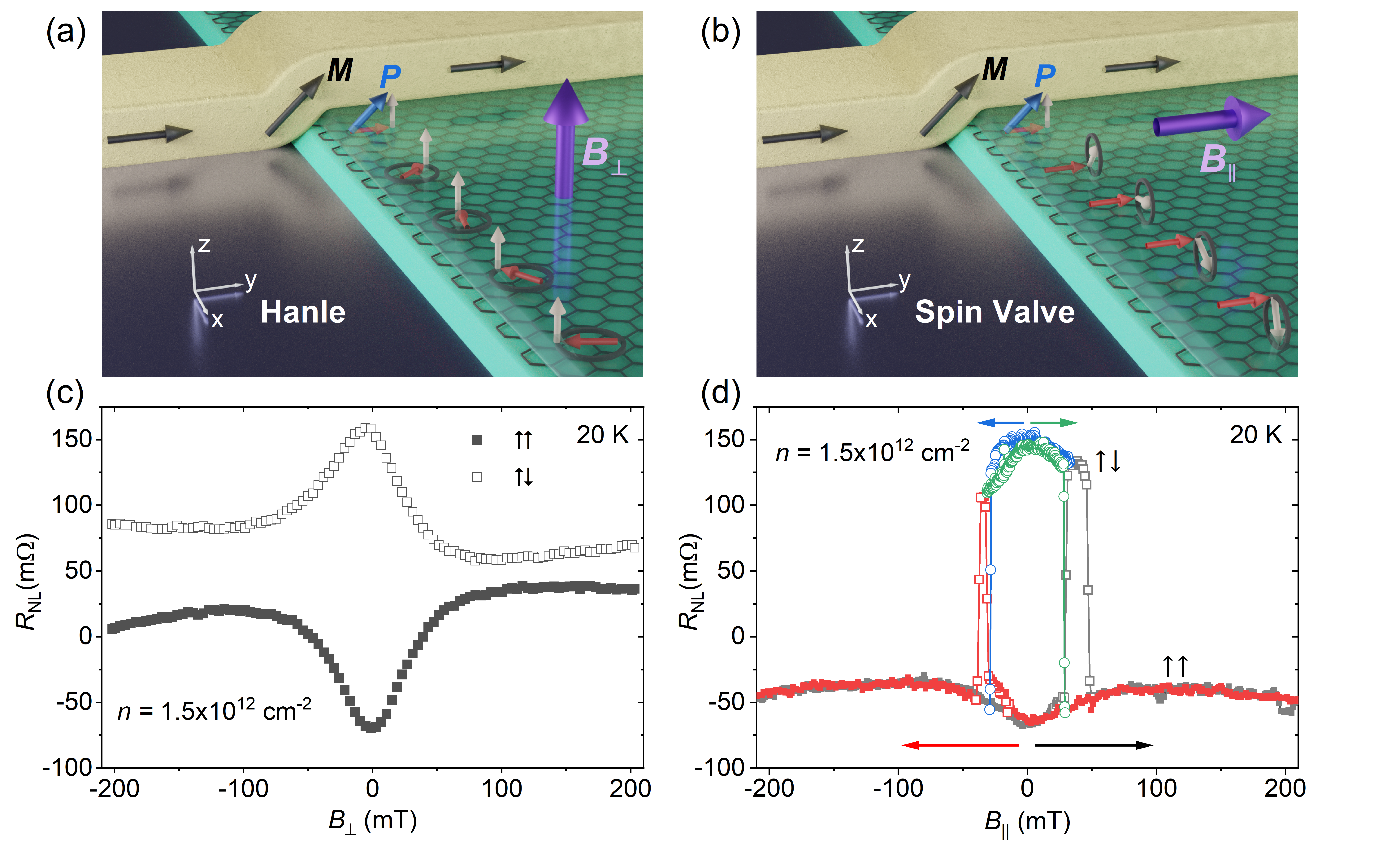}
\caption{ 
Defining Hanle and spin valve (SV) measurement configurations. \textbf{(a)} Illustration spin injection and Hanle measurement. The magnetic field, $B_\perp$ (purple arrow), points along the z-axis, causing the precession of the in-plane spin component, $P_{||}$ (red arrow). The total polarisation vector, $P$ (blue arrow), of the injected spin population follows the magnetisation vector of the nanowire, $M$ (black arrows). \textbf{(b)} Illustration of a SV measurement, for which the magnetic field, $B_{||}$ (purple arrow), is directed along the easy-axis of the nanowires (y-axis). Due to the geometry of these devices an out-of-plane spin component, $P_{\perp}$ (white arrow), is injected, which precesses around the $B_{||}$. \textbf{(c)} Hanle response for the cases of parallel ($\uparrow \uparrow$) and anti-parallel ($\uparrow \downarrow$) alignment of the magnetisation for the injector and detector electrodes.  \textbf{(d)} SV response of the device, with switches between the $\uparrow \uparrow$ and $\uparrow \downarrow$ states at $B_{||}\sim\pm40$ mT (indicated by empty black and red squares). A Hanle-like line shape is visible in the SV signal associated with the $\uparrow \uparrow$ state (indicated by filled red and black squares). When the magnetisation state switches to $\uparrow \downarrow$, the line shape is inverted, implying the effect is spin dependent (indicated by empty green and blue circles).
}
\label{fig:figure2}
\end{figure*}
    
We investigate the effect of the nanowire geometry on diffusive spin transport, by performing non-local (Figure \ref{fig:DevSchematic}a) spin precession (Hanle) and spin valve (SV) measurements, on device A \cite{Tombros-2007, Ingla-Aynes-2021}.  For Hanle measurements, a magnetic field, $B_{\perp}$, is applied along the z-axis, perpendicular to the plane of the channel, causing precession of the in-plane component of the spin polarisation, $P_{||}$, as the carriers diffuse through the channel (Figure \ref{fig:figure2}a). The observed Hanle response is characteristic of spin precession through the graphene channel (Figure \ref{fig:figure2}c). When ${B_{\perp}}=0$ there is a peak in the non-local signal, $R_{NL}$, as the spin polarisation is fully co-linear with the magnetisation of the electrode. As $B_{\perp}$ increases, $R_{NL}$ decreases to a minimum, at $B_{\perp} \sim \pm 100$ mT, implying a full $\pi$-rotation of the spin polarisation at the detector electrode. For $B_{\perp}> \pm100$ mT, $R_{NL}$ flattens out, suggesting the spin signal is fully dephased before reaching the detector. This effect is inverted for the two relative orientations of the injector-detector magnetisations: parallel ($\uparrow \uparrow$) and anti-parallel ($\uparrow\downarrow$) (Figure \ref{fig:figure2}a,c).

\begin{figure*}
    \includegraphics[width=\textwidth]{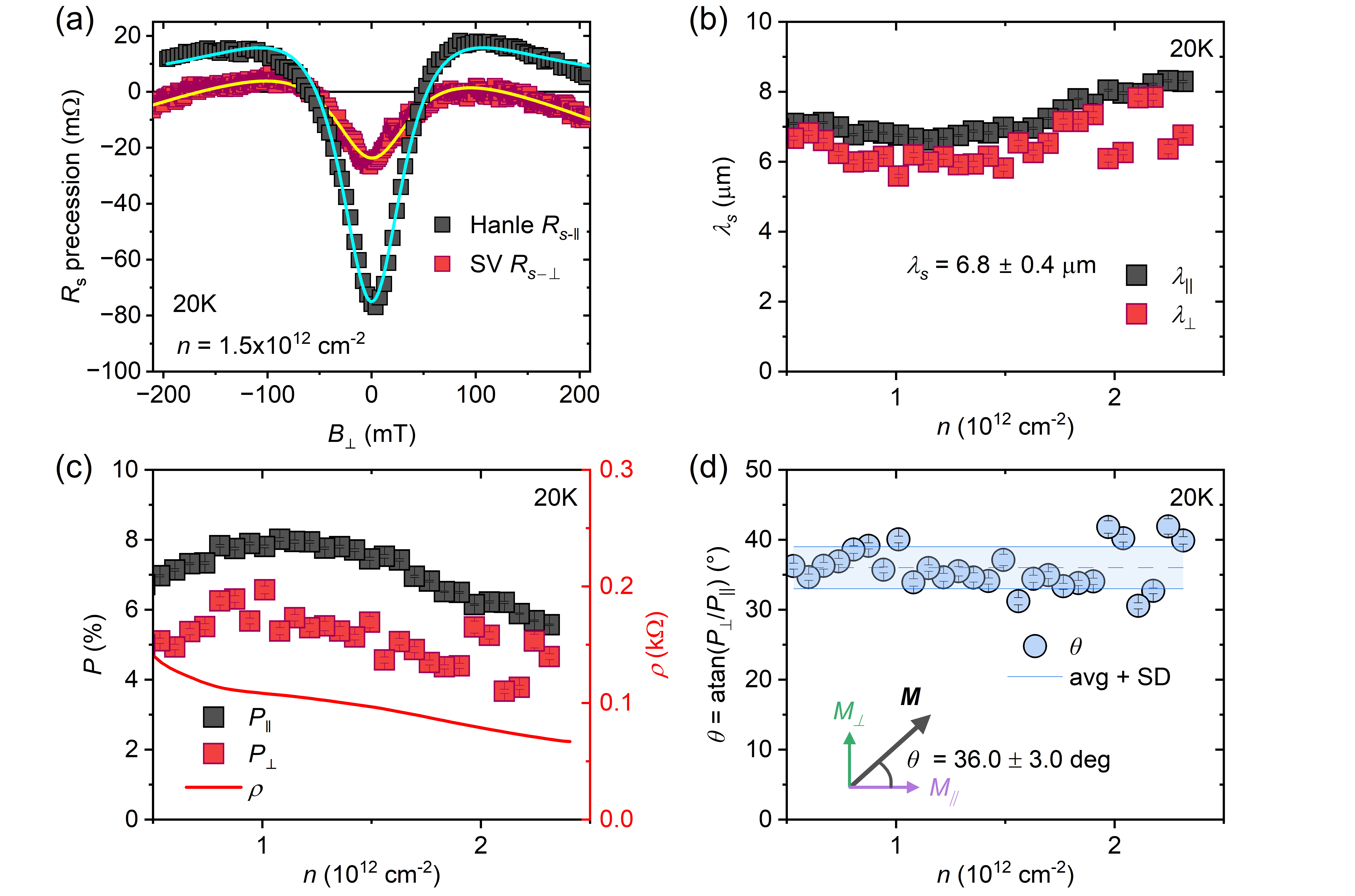}
    \caption{ 
Magnetisation angle reconstruction from spin transport \textbf{(a)} Comparison of spin-precession phenomena in terms of spin signal, $R_s$, at charge carrier density $n=\num{1.5E12}$ \si{\centi\metre^{-2}}. Black squares represent the full Hanle precession signal, while red squares represent the Hanle-like background from Figure \ref{fig:figure2}(d), with the $\uparrow\downarrow$ switches removed. Both data sets are presented with a linear background removed and offset corrections. A coloured line through each precession curve indicates a fit to the data. \textbf{(b)} Spin diffusion length, as a function of carrier density, extracted by fitting data from the two separate precession experiments over a range of carrier densities: Hanle ($\lambda_{||}$) and SV ($\lambda_\perp)$. The value and trend of $\lambda_s$ from the two precession phenomena are consistent across most of the carrier density range. \textbf{(c)} Spin polarisation as a function of carrier density, also extracted from fits to the two data sets; $P_{||}$ from Hanle data, and $P_\perp$ from SV data. We observe the same trend for both precession phenomena, indicating they describe the same system. The resistivity of the channel, $\rho$, is shown superimposed (right axis). \textbf{(d)} Inclination angle of the nanowire magnetisation, reconstructed from $\theta = \arctan (P_{\perp}/P_{||})$. The average angle across the data range is reported along with the standard deviation (shown as an error band). Error bars on data points in (b-d) come from individual fits to spin precession data as shown in (a).
    }
    \label{fig:figure3}
\end{figure*}

For spin valve measurements, the magnetic field, $B_{||}$, is applied along the y-axis, co-linear with the easy axis of the magnetic nanowires (Figure \ref{fig:figure2}b,d). The SV response seen here is unique to the geometry of our devices. The expected switches between the $\uparrow \uparrow$ and $\uparrow \downarrow$ states are observed at $B_{||} \sim \pm 40$ mT (open squares, Figure \ref{fig:figure2}d). However, the non-local signal associated with the $\uparrow \uparrow$ state, outside of the switching region, features a line shape reminiscent of that of a Hanle measurement (filled squares, Figure \ref{fig:figure2}d). We define this line shape as the Hanle-like background of the spin valve measurement. This phenomenon arises from the precession of an out-of-plane component in the spin polarisation, $P_{\perp}$, imprinted on the injected spin population by the magnetisation of the magnetic nanowire at the injection point. This component is also injected during Hanle measurements but will not precess in this case, as it is co-linear with the applied magnetic field. To determine whether the Hanle-like background is truly spin-dependent (i.e magnetic in origin), we perform a minor hysteresis loop, by reversing the direction of the magnetic field as soon as a switch occurs between the $\uparrow \uparrow$ and $\uparrow\downarrow$ states. The result of this process is shown in Figure \ref{fig:figure2}(d) by the open green and blue circles. The curve between the switches is inverted depending on whether the nanowire magnetisations are arranged parallel or antiparallel, as is the case for precession measurements, implying that the observed line shape is indeed spin-dependent. Therefore, providing additional evidence that the detected Hanle-like background arises from the precession of the out-of-plane component of injected spins.

\section{\label{sec:electronic_angle}Electronic angle reconstruction}

\begin{figure*}
\includegraphics[width=\textwidth]{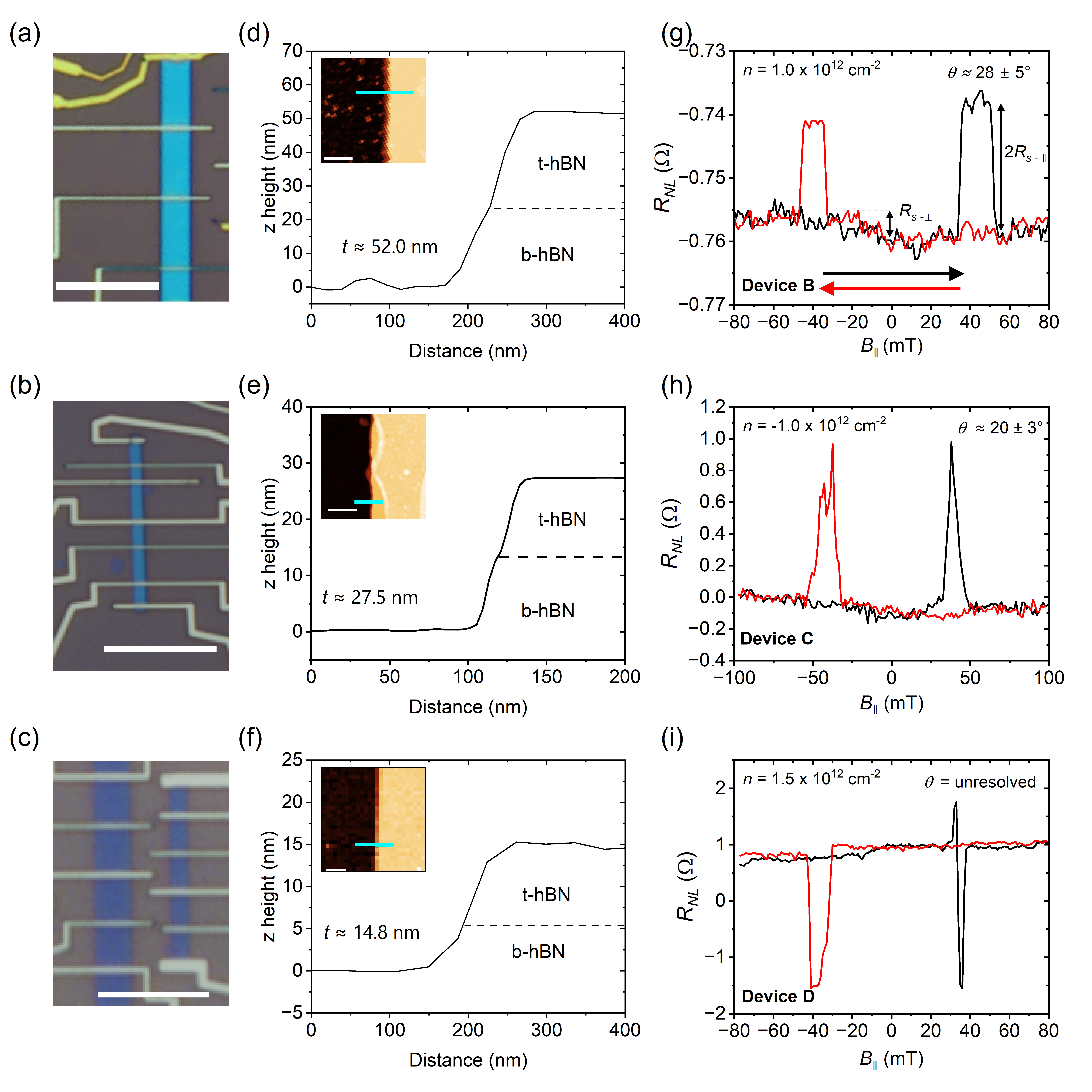}
\caption{Connection between heterostructure thickness, $t$, and observed Hanle-like background. \textbf{(a-c)} Optical microscope images of device B-D, arranged from thickest (B) to thinnest (D). The optical contrast can be used as a metric for stack thickness, with thinner flakes appearing darker in colour. White scale bars are 10 $\mu$m. \textbf{(d-f)} Height profiles taken from AFM scans at device edges. Stack thicknesses, from top to bottom, are:  $\sim 52.0$ nm (B), $\sim 27.5$ nm (C), and $\sim 14.8$ nm (D). Insets show the AFM images that profiles are extracted from, with the cyan bar illustrating profile locations (white scale bars are all 200 nm). \textbf{(g-i)} Spin valves taken from devices B (top), C (middle), and D (bottom), all at a temperature of 20 K. Red and black lines indicate the direction of the magnetic field sweep. A Hanle line shape is visible for device B. Magnitudes of in-plane and out-of-plane spin signals are indicated ($R_{s - ||}$ and $R_{s - \perp}$ respectively). Stack C displays a subtle Hanle-like background, while stack D does not show any. All spin valve measurements are taken at a carrier density far from the neutrality point.}
\label{fig:CompareThickness}
\end{figure*}

Figure \ref{fig:figure3}a shows representative spin precession data for two types of experiments: i) full Hanle signal, defined as $R_{s-||} = (R_{NL \uparrow\uparrow}-R_{NL \uparrow\downarrow})/2$, representing precession of a large in-plane spin component and ii) the Hanle-like background from the SV experiment, with the switches removed, representing precession of a small out-of-plane spin component ($R_{s-\perp}=R_{NL}$). We use a DC voltage, $V_{BG}$, applied between the silicon substrate, which acts as a global back gate (Figure \ref{fig:DevSchematic}a), and the graphene channel, to control the carrier density in graphene, $n$ \cite{Novoselov-2004, CastroNeto-2009}. The data is taken at a fixed charge carrier density where the non-local signal is maximum ($n=1.5\times10^{12}$ cm$^{-2}$; $V_{BG}=20$ V). A linear background is removed from both data sets, which arises due to Hall effect in the presence of a magnetic field. Both spin precession signals are fitted to the steady state solution to the Bloch equation, as has been discussed elsewhere \cite{Johnson1988, Tombros-2007, Tombros-2008, Swartz-2013, Ingla-Aynes-2021}. This process is repeated for data taken over a range of carrier densities ($n=0.5-2.5\times10^{12}$ cm$^{-2}$), which enables estimations of the spin relaxation time, $\tau_s$, spin diffusion coefficient, $D_s$, and contact polarisation, $P$. These values allow us to quantify the spin diffusion length, $\lambda_s = \sqrt{D_s\tau_s}$, the value of which is consistent across the two separate measurement techniques, with an average of $\lambda_s = 6.8 \pm 0.4$ $\mu$m (Figure \ref{fig:figure3}b).

The polarisation vector, $P$, is useful for studying the geometry of the device, as it is formed from orthogonal components of different magnitudes, $P_{\perp}$ and $P_{||}$. It relates to the non-local signal measured in the channel, $R_{NL}$, as \cite{Tombros-2007},

\begin{equation}
    R_{NL} = \frac{\rho P^2 \lambda_s}{2W}e^{-L/\lambda_s},
    \label{eq:RNL}
\end{equation}

\noindent
where we assume $P_i = P_d = P$ ($P_i$ and $P_d$ are injector and detector polarisation, respectively), $\rho$ is the resistivity of graphene, while $W$ and $L$ are the channel width and length, respectively. Hence, the observed difference in magnitude between the precession data extracted from SV and Hanle measurements, reflects the difference in the magnitude of the respective components of the polarisation, $P_{\perp}$ and $P_{||}$. Figure \ref{fig:figure3}c shows the evolution of these two components of polarisation against carrier density, with $P_{\perp}$ being consistently smaller than $
P_{||}$ throughout. At high density ($n\geq 2\times10^{12}$ cm$^{-2}$), the signal to noise ratio of $R_{NL}$ decreases, as it is proportional to $P^2$ and $\rho$ (Equation \ref{eq:RNL}, both of which decrease at high $|n|$ (Figure \ref{fig:figure3}c) \cite{Guarochico-Moreira-2022}. Hence, reliable fits are harder to perform as $n$ increases, especially for Hanle-like backgrounds, which are considerably smaller in magnitude than the full Hanle signal. So, only data up to $n=2\times10^{12}$ cm$^{-2}$ are used in the angle calculation. From the ratio $P_{\perp}/P_{||}$, we reconstruct the angle of the overall spin polarisation vector, $\theta = \arctan (P_{\perp}/P_{||})$. Note that this method yields an inclination angle which is insensitive to variations in the magnitude of the polarisation from contact to contact, being only sensitive to variations on its inclination. The extracted $\theta$ has an average value of $36.0\pm3.0$\si{\degree}, over the carrier density range explored (Figure \ref{fig:figure3}d). This agrees within experimental error with the angle of the nanowire extracted from AFM profilometry. Furthermore, the calculated value of $\theta$ remains constant with changing $n$, confirming that it reflects the geometry of the magnetic nanowire. Hence, we demonstrate geometric control over the spin polarisation injected into a graphene spin transistor.

We investigate the consistency of oblique spin injection across several devices of varying heterostructure thickness, all with a 60 nm thick Co layer in the nanowires. The right column of Figure \ref{fig:CompareThickness} shows spin valve signals measured from devices B, C, and D, with heterostructure thicknesses $t\approx 52.0$ nm, $t\approx27.5$ nm, and $t\approx14.8$ nm, respectively. In all cases, the magnitude of charge carrier density is comparable to that of the measurement on device A. Starting with the thickest of these stacks, device B (Figure \ref{fig:CompareThickness}c), a Hanle-like background is observable in the spin valve signal, with a magnitude $R_{s - \perp}\propto P_\perp$ (Equation \ref{eq:RNL}). Likewise, the magnitude of the switches is $2R_{s - ||}\propto P_{||}$ (Figure \ref{fig:CompareThickness}c). Hence, the ratio between these magnitudes can be used to approximate the inclination angle of the nanowire, which, for device B is found to be $\theta\approx28\pm5\degree$. Moving to the middle case, device C, the spin valve background shows a subtle line-shape, implying this thickness is a boundary for achieving detectable out-of-plane spin polarisation. The angle extracted here is $\theta\approx20\pm3\degree$. Finally, for the thinnest case, device D, the spin valve background is mostly flat. Here, the only measurable feature is a faint hysteresis loop around $B\sim-30$ mT, consistent with prior observation of Hall signal close to 1D magnetic contacts \cite{Karpiak-2017}. As such, the angle of the nanowire cannot be evaluated for this device, implying that any out-of-plane component in the magnetisation of the nanowire is below the resolution of the measurement. It is expected that using this approximate method should be less accurate for evaluating the nanowire geometry than the full Hanle analysis laid out above. Nevertheless, the angles follow the expected trend with decreasing thickness. Hence, the role of heterostructure thickness, relative to the Co layer thickness, is crucial in achieving a geometrically defined out-of-plane component in the nanowire magnetisation and, in turn, the spin polarisation.

\section{\label{sec:concluding_remarks} Conclusions}

In conclusion, we exploit spin transport in an encapsulated graphene spin valve device to characterise the magnetisation vector of a magnetic nanowire acting as a 1D edge contact. 
This relatively simple approach provides an alternative method of spin vector magnetometry that relies only on pure spin signal, enabling a quantitative analysis of a nanoscale magnetic region. Importantly, we demonstrate that the geometry of Van der Waals heterostructures can be engineered to define the angle of spin injection into 2D materials. The use of geometry to control the generation of out-of-plane spin is comparatively simpler than alternative methods based on unconventional spin-orbit effects \cite{Das-2018, safeer_spin_2020} and is generally applicable to any magnetic material without the need of magnetocrystalline anisotropy \cite{Zhao-2023}. Such structures can potentially be applied for the study of spin anisotropy in low-dimensional materials \cite{raes_determination_2016} via the use of a geometrically defined oblique spin accumulation.

\begin{acknowledgments}
UK participants in Horizon Europe Project “2D Heterostructure Non-volatile Spin Memory Technology” (2DSPIN-TECH) are supported by UKRI grant number [10101734] (The University of Manchester). We wish to acknowledge the support of Noel Natera-Cordero and Christopher R. Anderson for providing helpful insights towards the fabrication and characterisation of devices. D.B. acknowledges the Engineering and Physical Sciences Research Council (EPSRC), Doctoral Training Partnership (DTP) for funding the PhD project (U.K). J.C.T.F. acknowledges support from the Consejo Nacional de Ciencia y Tecnología (México). V.H.G.M. acknowledges support from the Secretaría Nacional de Educación Superior, Ciencia y Tecnología (SENESCYT), for the Ph.D. scholarship under the program “Universidades de Excelencia 2014” (Ecuador). Research data are available from the authors upon request.
\end{acknowledgments}

\bibliography{BIB}

\end{document}